\documentclass[twocolumn,preprintnumbers,amsmath,amssymb,superscriptaddress,prl]{revtex4}

\usepackage{graphicx}
\usepackage{epsfig}
\usepackage{color}
\usepackage{verbatim}
\usepackage{subfigure}
\usepackage{float}
\usepackage[greek,english]{babel} 
\usepackage[OT2,OT1]{fontenc} 
\begin{document}
%\maketitle
\preprint{APS/123-QED}

\title{A unified theory to describe the transition of stable nanobubbles to unstable microbubbles on homogeneous surface}% Force line breaks with \\
%\thanks{Corresponding authors.}%

\author{Binghai Wen}
\affiliation{Guangxi Key Lab of Multi-source Information Mining $\&$ Security, Guangxi Normal University, Guilin 541004, China.}

\author{Yongcai Pan}%
\affiliation{Guangxi Key Lab of Multi-source Information Mining $\&$ Security, Guangxi Normal University, Guilin 541004, China.}

\author{Lijuan Zhang}%
\affiliation{Shanghai Synchrotron Radiation Facility, Shanghai Advanced Research Institute, Shanghai 201204, China.}

\author{Shuo Wang}%
\affiliation{Julong college, Shenzhen Technology University, Shenzhen, 518118, China.}

\author{Limin Zhou}\thanks{Corresponding authors.\\zhoulimin@zjlab.org.cn; wangchunlei@zjlab.org.cn\\}
\affiliation{Shanghai Synchrotron Radiation Facility, Shanghai Advanced Research Institute, Shanghai 201204, China.}

\author{Chunlei Wang}\thanks{Corresponding authors.\\zhoulimin@zjlab.org.cn; wangchunlei@zjlab.org.cn\\}
\affiliation{Shanghai Synchrotron Radiation Facility, Shanghai Advanced Research Institute, Shanghai 201204, China.}

\author{Jun Hu}
\affiliation{Shanghai Synchrotron Radiation Facility, Shanghai Advanced Research Institute, Shanghai 201204, China.}


\begin{abstract}
Experiments have not only revealed the remarkably long lifetime of nanobubbles, but also demonstrated the diffusive instability of bubbles above micrometers, thus a full-scale physical understanding on the stability of bubbles is in urgent need. Herein, we develop a model that captures the state transition from stable nanobubbles to unstable microbubbles on homogeneous surfaces. The transition explains the typical long lifetime, limited height and small contact angle of surface nanobubbles observed in experiments. The consequent phase diagram shows that the bubble size and gas dissolving saturation determine the dynamic behaviors of surface bubbles, namely growth, stability, shrinkage or dissolution.
\end{abstract}

\maketitle

%\tableofcontents

Surface nanobubbles, which are spherical-cap gaseous domains with nanoscale thickness attached on immersed substrates, exhibit surprising long lifetime surviving up to days and weeks \cite{RN5080, RN5308, RN5309, RN5317, RN5416, RN5433}. They have a special flat morphology; typically, the heights are less than 100 nm, and the gas-side contact angles are around ${20}^{\circ}$ and appear to be weakly dependent on substrate wettability \cite{RN5422, RN5082, RN5381, RN5424, RN5375, RN5382, RN5123, RN5466}. The remarkable stability has perplexed researchers for nearly two decades, because it contradicts the widely accepted Epstein-Plesset theory, which was established for bulk bubbles and then adopted for surface bubbles described by the curvature radius \cite{RN5315, RN5080, RN5416}. This theory suggests that all gas bubbles are diffusively unstable in practical situations, experiencing either shrinkage to dissolution or unbounded growth \cite{RN5448, RN5122, RN5416}. Notably, the theoretical prediction is in excellent agreement with experiments on bubble radii down to micrometers \cite{RN5456, RN5416}. However, for nano-sized bubbles, it gives the lifetime corresponding to the diffusion timescale $ \tau \sim{R_{0}^{2}{{\rho }_{g}}}/{(D{{c}_{s}})} $ and suggests that a typical nitrogen nanobubble with the initial radius $ {{R}_{0}}= $ 100 nm must dissolve within 100 ${\mu s}$, where ${{\rho }_{g}}$ is the gas density, ${{c}_{s}}$ is the gas solubility and \textit{D} is the diffusion constant for the gas in the liquid \cite{RN5315, RN5080}. Thus, there is an imminent need to theoretically unify the remarkable stability of nanobubbles and the general instability of microbubbles.

The stability of surface bubbles requires gas diffusive equilibrium at the liquid-bubble interface to maintain the volume at constant, meanwhile it also requires mechanical equilibrium at the three-phase contact line to keep the bubble immobile. Contact line pinning was considered to be a prerequisite to stabilize surface nanobubbles \cite{RN5119, RN5310, RN5118}. Besides settling the mechanical equilibrium readily, contact line pinning compels a nanobubble to reduce the curvature and Laplace pressure when its volume decreases, and then promotes its stability \cite{RN5363}. But pinning cannot work alone to prevent the bubble's shrinkage and dissolution, because the Laplace pressure tends to press any gas out of bubbles \cite{RN5106,RN5313, RN5080}. Gas oversaturation of the ambient liquid effectively counteracts the gas leakage from the nanobubble, even though it has already been reduced to a small amount by the pinning effect \cite{RN5447, RN5313, RN5106}. Nevertheless, experiments indicated that nanobubbles could survive in an open or degassed liquid \cite{RN5318, RN5209, RN5118}. An attractive potential was introduced to enrich the dissolving gas adjacent to hydrophobic substrate and stabilize nanobubbles in undersaturated environments \cite{RN5121, RN5427, RN5085, RN5512, RN5420}. Then, by considering gas transport as the bulk liquid equilibrated with the external environment, the stability and dynamics of surface nanobubbles were consistent with experimental response timescale \cite{RN5120, RN5119}. Although contact line pinning together with oversaturation or hydrophobic potential could effectively stabilized surface nanobubbles, experiments have demonstrated that nanobubbles could be readily moved on PFOTS, PDMS or polymer brushes surfaces by using an AFM tip \cite{RN5102, RN5212, RN5123}. These phenomena are in agreement with the simulations, which also supported that contact line pinning was not strictly required for the mechanical equilibrium of surface nanobubbles \cite{RN5323, RN5210}. Recently, the adsorption of gas molecules at the substrate beneath surface nanobubbles was taken into account to lower the energy of the solid-gas interface and explain the flat nanobubble morphology \cite{RN5086}. But the thermodynamic model has ignored the diffusive equilibrium by assuming the liquid-bubble interface would be impermeable.

\graphicspath{ {Figures/} }

\begin{figure}[b]
	\includegraphics[width=8.6cm]{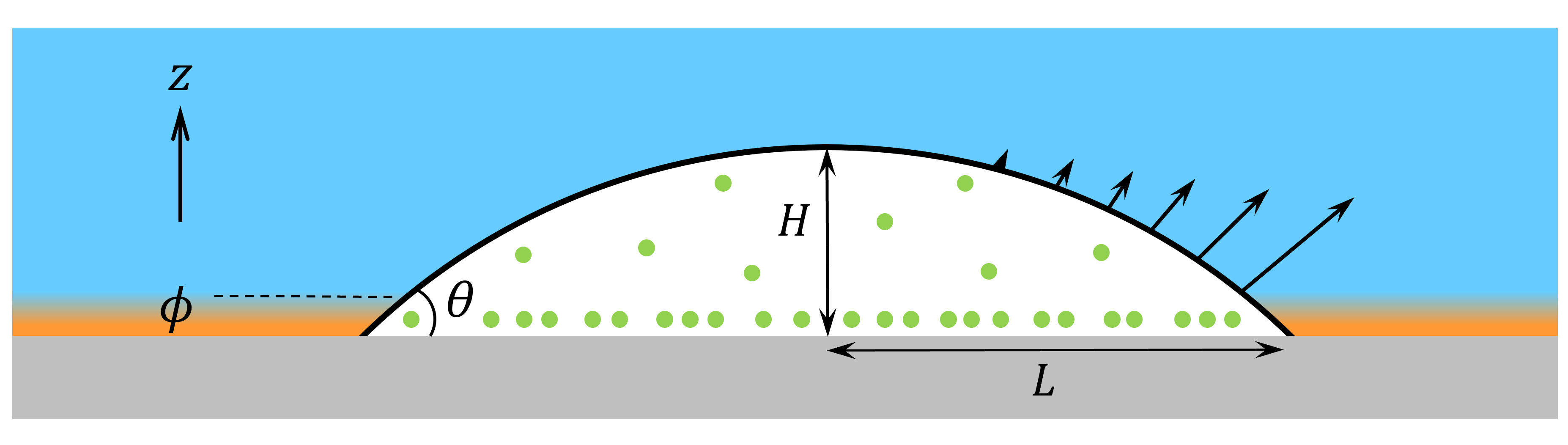}% Here is how to import EPS art
	\caption{\label{fig:epsart} Schematic diagram of a surface nanobubble.}
\end{figure}

In this letter, we propose a full-scale model that captures bubble dynamics at homogeneous substrate, which can simultaneously achieve both the diffusive and mechanical equilibriums, and uniformly illustrate the nanobubble stability and the instability of larger bubbles above microscale.

Fig. 1 presents the schematic diagram of a surface nanobubble with a spherical cap shape, which can be parameterized by the footprint radius \textit{L}, the bubble height \textit{H} and the gas-side contact angle ${\theta}$. The substrate is plane and homogeneous precluding softness, physical roughness or chemical heterogeneity \cite{RN5416}. The Laplace equation gives the pressure inside the nanobubble $P={{P}_{0}}+{2{{\gamma }_{\lg }}}/{R}$, where ${{P}_{0}}$ is atmospheric pressure, \textit{R} is the curvature radius, and ${{\gamma }_{\lg }}$ is the liquid-gas surface tension. The gas concentration at the liquid-bubble interface can be given by Henry's law $c(t)\text{=}{{{c}_{s}}P(t)}/{{{P}_{0}}}$, where ${{c}_{s}}$ is the gas solubility \cite{RN5106}. 

The gas diffusion from a surface nanobubble is analogous to the evaporation of a pinned drop \cite{RN2934}, which was exactly solved by Popov \cite{RN3044}. Incorporating Henry's law into Popov's solution, Lohse and Zhang derived the mass change of a pinned nanobubble in a supersaturated liquid \cite{RN5106},
%Eq.(1)
\begin{equation}
\frac{dM}{dt}=-\pi LD{{c}_{s}}f(\theta )\left( \frac{2{{\gamma }_{\text{lg}}}}{L{{P}_{0}}}\sin \theta -\zeta  \right),\\
\end{equation}
\noindent where ${\zeta ={c}/{{{c}_{s}}}-1}$ is the gas oversaturation of the environmental liquid, and the geometric term depends on the gas-side contact angle,

\begin{equation}
f(\theta )\text{=}\frac{\sin \theta }{1\text{+}\cos \theta }+4\int_{0}^{\infty }{\frac{1+\cosh 2\theta \xi }{\sinh 2\pi \xi }\tanh \left[ (\pi -\theta )\xi  \right]d\xi}.\\
\end{equation}

Once the diffusive equilibrium appears, namely ${{dM}/{dt}\;=0}$, the contact angle is dependent on the oversaturation by ${\sin \theta ={L{{P}_{0}}\zeta }/{2{{\gamma }_{\lg }}}}$. Tan \textit{et al.} \cite{RN5121} further explained the experimental observations that surface nanobubbles can survive in undersaturated environments by introducing a short-ranged attraction potential ${{\phi }_{\text{0}}}$ on hydrophobic substrates \cite{RN5318, RN5321, RN5209, RN5420}. The attraction potential induced a supersaturated gas reservoir even if the bulk liquid is undersaturated and then changed the distribution of dissolved gas oversaturation adjacent to the substrate \cite{RN5121}, 

\begin{equation}
\zeta (z)=\frac{{{c}_{\infty }}}{{{c}_{s}}}\exp \left( -\frac{{{\phi }_{\text{0}}}{{e}^{-z\text{/}\lambda }}}{{{k}_{B}}T} \right)-1,\\
\end{equation}

\noindent where ${\lambda \sim}$1 nm is the characteristic distance of the interaction. This supersaturated layer compensates for the gas leakage in the upper part of nanobubbles, because they dynamically respond to the local changes of the gas concentration in the surrounding liquid \cite{RN5456, RN5121}. 

For a surface nanobubble, pinning is helpful for its stabilization instead of a prerequisite \cite{RN5102, RN5212, RN5323, RN5210}. In fact, the Popov's equation does not strictly require contact line pinning \cite{RN3044}, which is just necessary to form a stain ring during drop evaporation \cite{RN2934}. Hence the pinning model developed by Lohse, Zhang and Tan \textit{et al.} \cite{RN5106, RN5121} can be adopted to calculate the dynamic diffusion of a nanobubble with a variable footprint radius, but the mechanical equilibrium must be settled firstly. 

Essentially, the mechanical equilibrium at an unpinned contact line is illuminated by the Young's equation, which relates the interface energies to the contact angle \cite{RN5101, RN5467}. Petsev \textit{et al.} considered that the adsorption of gas molecules to the substrate lowered the solid-gas interface energy and explained the flat morphology of nanobubbles from a thermodynamic perspective \cite{RN5086}. The adsorption effect is described by the Langmuir adsorption isotherm, which is thermodynamically equivalent to the Szyszkowski equation \cite{RN5086, RN5478},

\begin{equation}
\gamma _{\text{sg}}^{\text{*}}\text{(}P\text{)}\text{=}\gamma _{\text{sg}}^{0}-\frac{{{k}_{B}}T}{b}\ln (1+KP),\\
\end{equation}

\noindent where \textit{K} is the equilibrium adsorption constant, \textit{b} is the cross sectional area of a single adsorbing molecule, $\gamma _{\text{sg}}^{0}$ is the surface tension of the solid-gas interface in a vacuum or absence of adsorption. Combining Eq. (4) and the Young's equation $\gamma _{\text{sg}}^{\text{*}}-{{\gamma }_{\text{sl}}}\text{+}{{\gamma }_{\text{lg}}}\cos \theta  \text{=} \text{0}$ gives a general expression for the influence of gas adsorption \cite{RN5086}, 

\begin{equation}
\gamma _{\text{sg}}^{0}-{{\gamma }_{\text{sl}}}\text{+}{{\gamma }_{\text{lg}}}\cos \theta  - \frac{{{k}_{B}}T}{b}\ln (1+KP)=0,\\
\end{equation}

\noindent where ${{\gamma }_{\text{sl}}}$ is the solid-liquid surface tension. However, this adsorption model does not settle the diffusive equilibrium instead of assuming no transfer of gas molecules across the liquid-bubble interface \cite{RN5086}.

\begin{figure*}[t]
	\includegraphics[width=15cm]{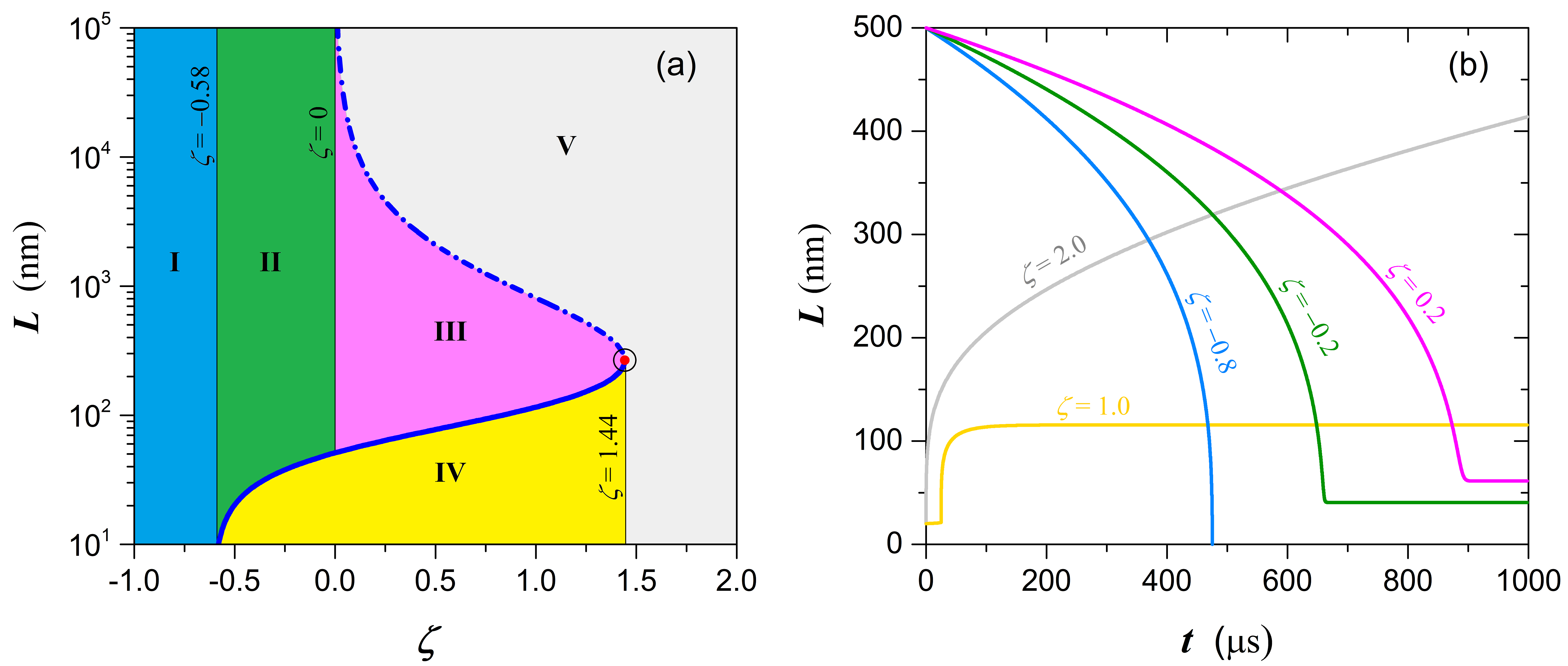}% Here is how to import EPS art
	\caption{\label{fig:wide}(a) The phase diagram of bubble dynamics on a homogeneous substrate with ${{\theta }_{0}}={{{60}^{\circ}}}$. The inflection point at ${\zeta}$= 1.44 divides the curve into the solid and dashdot parts, which indicate the stable and unstable equilibrium states of surface bubbles, respectively. The curve together with the lines ${\zeta}$= -0.58, 0 and 1.44 partition the diagram into five sections. (b) The dynamic evolutions of five typical nanobubbles in each section. The colors of the lines represent the sections they belong to. The evolution of the nanobubble in the section IV (yellow) has been postponed by 25 ${\mu}$s in order to avoid data overlap.}
\end{figure*}

Obviously, the pinning and adsorption models are exactly complementary when dealing with the diffusive and mechanical equilibriums, respectively. Molecular dynamics simulations showed that the timescale that thermal motion of molecules equilibrates a contact angle is within nanoseconds \cite{RN5210, RN2082, RN5206}, which is more than three orders of magnitude faster than the diffusion timescale. From the perspective of diffusion, surface nanobubbles always retain the mechanical equilibrium during dynamic growth or shrinkage. Substituting the solid-gas surface tension at atmospheric pressure ${{\gamma }_{\text{sg}}}\text{=}\gamma _{\text{sg}}^{\text{*}}\text{(}{{P}_{0}}\text{)}$ into Eq. (5) gives the contact angle with the mechanical equilibrium,

\begin{equation}
\cos \theta \text{=}\cos {{\theta }_{0}}\text{+}\frac{{{k}_{B}}T}{b{{\gamma }_{\text{lg}}}}\ln \left( \frac{1+KP}{1+K{{P}_{0}}} \right),\\
\end{equation}

\noindent where ${{\theta }_{0}}$ indicates the wettability of the substrate at atmospheric pressure, which is commonly reported in experimental literature. Replacing the contact line pinning by the adsorption effect, namely constraining the real-time contact angle by Eq. (6), we can derive the dynamic equation to evolve a bubble on a homogeneous surface

%Eq.(7)
\begin{widetext}
\begin{equation}
%\begin{align}
  \frac{dM}{dt}=-\frac{\pi LD{{c}_{s}}}{H}f(\theta ) \\ 
  \left[ \frac{2{{\gamma }_{\text{lg}}}}{{{P}_{0}}}\left( 1-\cos {{\theta }_{0}}-\frac{{{k}_{B}}T}{b{{\gamma }_{\text{lg}}}}\ln \frac{1+KP}{1+K{{P}_{0}}} \right)-\int_{0}^{H}{\zeta (z)dz} \right]. \\ 
%\end{align}
\end{equation}
\end{widetext}

When the evolution achieves ${dM}/{dt}\;=0$, this derives the equilibrium equation, 

\begin{equation}
\cos {{\theta }_{e}} \equiv  1-\frac{{{P}_{0}}}{2{{\gamma }_{\text{lg}}}}\int_{0}^{H}{\zeta (z)dz}=\cos {{\theta }_{0}}\text{+}\frac{{{k}_{B}}T}{b{{\gamma }_{\text{lg}}}}\ln \left( \frac{1+KP}{1+K{{P}_{0}}} \right),\\
\end{equation}

\noindent which ensures the diffusive and mechanical equilibriums simultaneously. Furthermore, it demonstrates that nanobubbles can be stabilized on a homogeneous surface, as long as the contact angle from diffusive equilibrium is equal to that required by mechanical equilibrium, without the help of pinning, softness, roughness or heterogeneity.

We consider a typical case of a bubble system at ambient conditions as illustrated in Fig. 1. The parameters are referred to the literature by Lohse and Zhang \cite{RN5106}, Tan \textit{et al.} \cite{RN5121} and Petsev \textit{et al.} \cite{RN5086}, viz. $D=2\times {{10}^{-9}}\text{ }{{m}^{2}}/s$, ${{c}_{s}}=0.017\text{ }kg/{{m}^{3}}$, ${{\rho }_{g}}=1.165\text{ }kg/{{m}^{3}}$, $\lambda =1\text{ }nm$. The cross-sectional area of the adsorbing gas molecule is found from the van der Waals diameter of nitrogen, $b=7.548\rm{\AA}^{2}$ \cite{RN5086}. Both the contact angle ${{\theta }_{0}}$ and the attractive potential ${{\phi }_{0}}$ relate to the wettability of the substrate, and the analysis shows that the relationship between them is approximately linear \cite{RN5420}. Thus, we use a linear connection here; a neutral substrate, which is equal to that in Lohse and Zhang model \cite{RN5106, RN5121}, has a zero potential and ${\theta }_{0}={60}^{\circ}$ corresponds to $\phi =-2$, where $\phi \text{ = }{{{\phi }_{\text{0}}}}/{({{k}_{B}}T)}$. The adsorption constants of different substrates span a broad spectrum of magnitudes, and \textit{K} is typically between $1.0\times {{10}^{-6}}$ $\sim$ $2.0\times {{10}^{-3}}\text{ }P{{a}^{-1}}$ for $N_2$, $O_2$, and $CO_2$ adsorbing to HOPG, MG-MOF or $Fe_2$ \cite{RN5086}. We choose a representative value of $K=1.0\times {{10}^{-5}}\text{ }P{{a}^{-1}}$ for qualitative experimental comparisons.

\begin{figure*}[htb]
	\includegraphics[width=15cm]{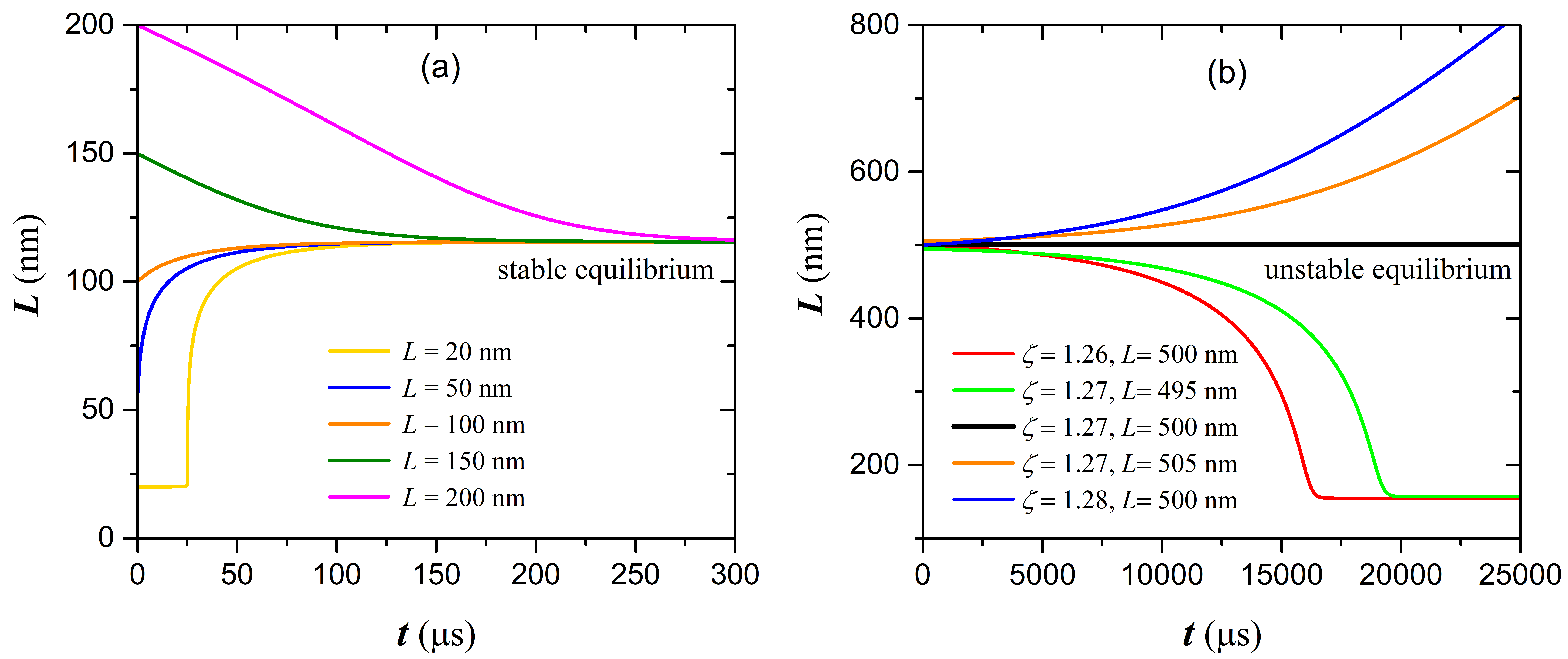}% Here is how to import EPS art
	\caption{\label{fig:wide}The comparison of the dynamic evolutions of nanobubbles around the stable and unstable equilibriums. (a) The dynamic evolutions of the five nanobubbles on both sides of the solid curve. The oversaturation takes ${\zeta}$ = 1 and the initial footprint radii are \textit{L} = 20, 50, 100, 150 and 200 nm, respectively. All of them converge to the point \textit{L} = 116 nm on the solid curve and achieve a stable equilibrium state. (b) The dynamic evolutions of the five nanobubbles on or adjacent to the dashdot curve. The central nanobubble with \textit{L} = 500 and ${\zeta}$ = 1.27 is on the dashdot curve, while the others around it have less than 1${\%}$ difference in the footprint radius or oversaturation. Since all of the around nanobubbles deviate from it, the central nanobubble represents an unstable equilibrium state.}
\end{figure*}

The dynamic equation (7) and equilibrium equation (8) depict the phase diagram of bubble dynamics on a homogeneous surface, as shown in Fig. 2. We solve Eq. (8) for the footprint radius as a function of the oversaturation on the substrate with ${{\theta }_{0}}={{{60}^{\circ}}}$. The functional curve begins at \textit{L} = 10 nm and ${\zeta}$ = -0.58. Since the corresponding height has already been less than 1 nm, the gaseous domain could be viewed as a gas nucleation. The curve has an apparent inflection point at ${\zeta}$ = 1.44 and \textit{L} = 267 nm, which separates the curve into the lower solid and upper dashdot parts. Above the inflection point, the dashdot curve gradually approaches the line ${\zeta}$ = 0 as the bubble grows to more than microns, because a large bubble has a low Laplace pressure and then the adsorption effect becomes weaker and weaker. The function curve together with the three lines ${\zeta}$ = -0.58, 0 and 1.44 partition the phase diagram into five sections in Fig. 2(a). Then, we solve Eq. (7) to investigate the bubble dynamics in each section, and five typical nanobubbles are selected correspondingly. The initial footprint radii are \textit{L} = 500 nm for the section I ${\sim}$ III and 20 nm for the section IV ${\sim}$ V. During the dynamic evolutions, the oversaturation of the bulk liquid keeps ${\zeta}$ = -0.8, -0.2, 0.2, 1.0 and 2.0, respectively. Fig. 2(b) presents that the nanobubble in the section I (blue) shrinks until it dissolves completely, and this process is accelerative since the Laplace pressure diverges as the curvature radius decreases. The nanobubbles in the section II (green) and III (magenta) also shrink, but they finally reach the stable state locating at the solid curve; notwithstanding the section III has an upper limit given by the dashdot curve. The nanobubble in the section IV (yellow) grows to the stable state at the solid curve, while that in the section V (gray) presents an unbounded growth up to microscale. The evolutions with some similar initial parameters were performed in the previous studies with the contact line pinning \cite{RN5106, RN5121}. The most typical difference caused by pinning is that the equilibrium height and contact angle are dependent on the constant footprint radius \cite{RN5106, RN5121}. The pinning nanobubbles reach the stable state faster than those in the present study; because, when the adsorption effect is considered, the inner pressure is lower, and then the gas diffusion across the liquid-bubble interface is slowed down \cite{RN5086}.

The phase diagram manifests that the dynamics of nanobubbles is determined by the size and oversaturation, and exhibits the behaviors of growth, stability, shrinkage or dissolution. The phase diagrams of ${{\theta }_{0}}={{{50}^{\circ}}}$ and ${{{70}^{\circ}}}$ are drawn in the supplemental material for comparisons. The substrate hydrophobicity significantly influences the division of the sections. A more hydrophobic surface can stabilize nanobubbles in a lower oversaturation, so the curve moves to the left. When the substrate is less hydrophobic, the larger oversaturation is required; thus, the section I expands and swallows the section II gradually. However, bubbles above microscale always approach the line ${\zeta}$ = 0, and this indicates that the influence of surface hydrophobicity is weak on large bubbles.

\begin{figure*}
\includegraphics[width=15cm]{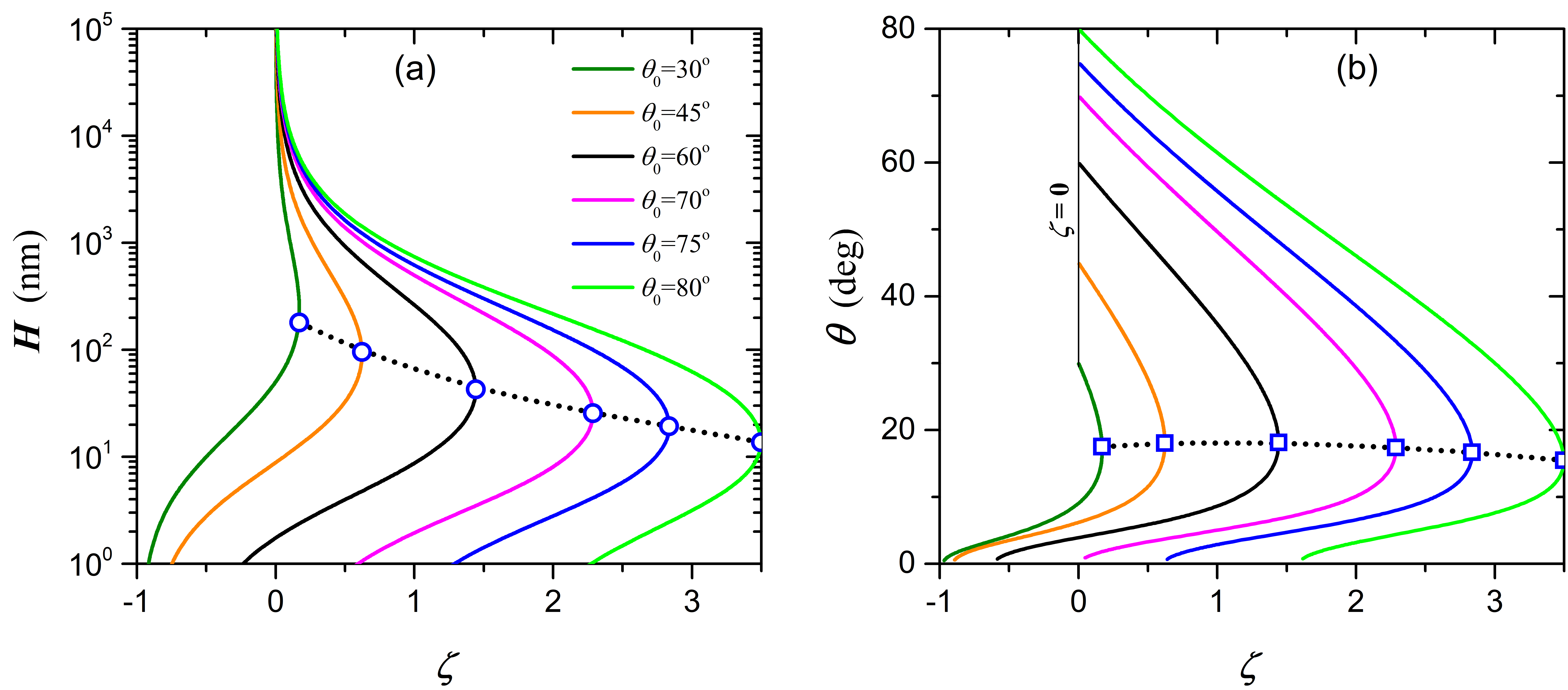}% Here is how to import EPS art
\caption{\label{fig:wide} (a) The height and (b) the contact angle of equilibrium bubbles as functions of the oversaturation on the substrates with various wettability. The inflection points of the bubble height and contact angle are marked by the circles in (a) and squares in (b), respectively.}
\end{figure*}

Next, we distinguish the stable and unstable equilibrium states of nanobubbles indicated by the lower solid curve and the upper dashdot curve, respectively. Two sets of evolutions are performed to investigate the bubble dynamics around the curve. For the solid part, five nanobubbles are selected with the oversaturation ${\zeta}$ = 1 and the initial footprint radius \textit{L} = 20, 50, 100, 150 and 200 nm, respectively. Fig. 3(a) presents that all the five cases evolve and reach the equilibrium point \textit{L} = 116 nm on the solid curve, while the relevant height and contact angle are \textit{H} = 9 nm and ${\theta} = {{{8.63}^{\circ}}}$. Importantly, the nanobubbles on both sides must converge to the solid curve; that is to say, even with some perturbations, they will quickly return to the stable state. Therefore, the solid curve represents the stable equilibrium of nanobubbles on a homogeneous substrate. For the dashdot curve in the phase diagram, we select the nanobubble with \textit{L} = 500 nm and ${\zeta}$ = 1.27 on the curve and the other four nanobubbles surrounding it. Fig. 3(b) presents that the nanobubbles on both sides always deviate from the dashdot curve, whereas the nanobubble on the curve keeps unchanged. Since the four surrounding nanobubbles have less than 1${\%}$ difference in the footprint radius or oversaturation relative to the central one, the equilibrium state represented by the central one can be easily broken by a slight perturbation from the bubble size or gas concentration. Therefore, the dashdot curve represents an unstable equilibrium of nanobubbles on a homogeneous substrate. 

These analyses demonstrate that the present theory uniformly describes the stability of nanobubbles, which has been confirmed by experiments \cite{RN5209, RN5082, RN5118, RN5115}, and the instability of bubbles with larger sizes up to microscale, which has been predicted by the Epstein-Plesset theory and validated by microscope observations \cite{RN5456, RN5448, RN5122, RN5416}. The section III connects the solid and dashdot curves simultaneously, thus, for a given oversaturation, bubbles on a homogeneous surface can be in the stable or unstable equilibrium state; the former corresponds to the free energy minimum, while the latter suggests the free energy maximum of bubble system \cite{RN5363, RN5214}. It is the inflection point that indicates the state transition of surface bubbles from the stable equilibrium to the unstable equilibrium. Except the equilibriums represented by the curve, nanobubbles at the rest of the phase diagram are in a nonequilibrium state, and their stabilizations in situ must get help from the contact line pinning.

Finally, we investigate the influence of surface wettability on the nanobubble stability. Fig. 4(a) presents the heights of equilibrium bubbles as functions of the oversaturation. With the decrease of surface hydrophobicity indicated by ${{{\theta}_{0}}}$, the curves of stable equilibrium become more and more to the right. This suggests that the larger oversaturation is required to equilibrate nanobubbles on lower hydrophobic substrates. Above the inflection points, as bubbles grow gradually up to microscale, the oversaturation required by the unstable equilibrium on every substrate approaches ${\zeta}$ = 0. This is consistent with the Epstein-Plesset equation and experiments, which confirmed that, above micrometers, bubbles could only be equilibrated in the solvent with gas saturation concentration \cite{RN5456, RN5448, RN5416}. Meanwhile, all the bubble contact angles touch the relevant contact angles ${{{\theta }_{0}}}$ of the substrates, as shown in Fig. 4(b). These reflect the decreasing of the Laplace pressure and the weakening of the adsorption effect in larger bubbles. Furthermore, the inflection points give the height and contact angle maximums of stable equilibrium nanobubbles on these homogeneous substrates. Fig. 4(a) shows that the critical heights are less than 100 nm for the hydrophobic substrates besides ${{\theta }_{0}}={{{30}^{\circ}}}$. Fig. 4(b) shows that all the critical contact angles are close to ${{20}^{\circ}}$. These results agree with the typical ranges of height and contact angle of nanobubble, and explain, in a certain degree, the weak dependence of the nanobubble contact angle on the substrate wettability observed in experiments \cite{RN5375, RN5382, RN5123, RN5422, RN5424, RN5082, RN5381, RN5466}.

In summary, we propose a model to evaluate the dynamics and stability of surface nanobubbles by integrating the gas diffusion mechanism, the hydrophobic attraction potential and the gas adsorption effect. It uniformly describes the state transition of the stable nanobubble to unstable microbubbles on homogeneous substrates. The dynamic evolutions show that the bubble size and environmental oversaturation determine the nanobubble behaviors, namely growth, stability, shrinkage or dissolution. Except the stable equilibrium nanobubbles, any other surface bubbles are in unstable equilibrium or nonequilibrium state on a homogeneous surface. Their stabilization has three routes: shrinking (in the sections II and III) or growing (in the section IV) to the stable equilibrium, otherwise relying on contact line pinning. Thus, contact line pinning can greatly expand the stable range of surface bubbles. The dynamic calculations in this letter only treat gas transport in the vicinity of the nanobubble; when gas transport is considered to equilibrate the bulk liquid with an open environment, the dynamics of surface nanobubbles can achieve to be consistent with experimental response timescale \cite{RN5106, RN5121, RN5120, RN5119}. The verification of the nanobubble stability on homogeneous surfaces appeals to rigorous experiments that control softness, physical roughness and chemical heterogeneity \cite{RN5416}.

\begin{acknowledgments}
This work was supported by the National Natural Science Foundation of China (Nos. 11862003, 12005284, 12022508 and 12074394), the Key Project of Guangxi Natural Science Foundation (No. 2017GXNSFDA198038), the Key Research Program of the Chinese Academy of Sciences (QYZDJ-SSW-SLH019).
\end{acknowledgments}

\end{document}